\documentclass[manuscript]{aastex}

\usepackage{graphicx}
\usepackage{xcolor}

\usepackage[normalem]{ulem}

\shorttitle{Photometric analysis of four overcontact binaries}
\shortauthors{Latkovi{\' c} et al.}

\begin{document}

\title{Photometric analysis of overcontact binaries\\
	AK Her, HI Dra, V1128 Tau and V2612 Oph}

\author{\c{S}. \c{C}al{\i}\c{s}kan}
\affil{Department of Astronomy \& Space Sciences, Faculty of Science,
  Ankara University, TR-06100 Tandogan, Ankara, Turkey}
\email{seyma.caliskan@science.ankara.edu.tr}

\author{O. Latkovi{\' c}}
\affil{Astronomical Observatory, Volgina 7, 11060 Belgrade, Serbia}
\email{olivia@aob.rs}

\author{G. Djura{\v s}evi{\' c}\altaffilmark{1}}
\affil{Astronomical Observatory, Volgina 7, 11060 Belgrade, Serbia}
\altaffiltext{1}{Isaac Newton Institute of Chile, Yugoslavia Branch}
\email{gdjurasevic@aob.rs}

\author{\.{I}. \"{O}zavc{\i}}
\affil{Department of Astronomy \& Space Sciences, Faculty of Science,
  Ankara University, TR-06100 Tandogan, Ankara, Turkey}
\email{ozavci@science.ankara.edu.tr}

\author{\"{O}. Ba\c{s}t\"{u}rk}
\affil{Department of Astronomy \& Space Sciences, Faculty of Science,
  Ankara University, TR-06100 Tandogan, Ankara, Turkey}
\email{obasturk@ankara.edu.tr}

\author{A. Cs{\'e}ki}
\affil{Astronomical Observatory, Volgina 7, 11060 Belgrade, Serbia}
\email{attila@aob.rs}

\author{H. V. \c{S}enavc{\i}}
\affil{Department of Astronomy \& Space Sciences, Faculty of Science,
  Ankara University, TR-06100 Tandogan, Ankara, Turkey}
\email{hvsenavci@ankara.edu.tr}

\author{T. K{\i}l{\i}\c{c}o\u{g}lu}
\affil{Department of Astronomy \& Space Sciences, Faculty of Science,
  Ankara University, TR-06100 Tandogan, Ankara, Turkey}
\email{tkilicoglu@ankara.edu.tr}

\author{M. Y{\i}lmaz}
\affil{Department of Astronomy \& Space Sciences, Faculty of Science,
  Ankara University, TR-06100 Tandogan, Ankara, Turkey}
\email{mesutyilmaz@ankara.edu.tr}

\and

\author{S. O. Selam}
\affil{Department of Astronomy \& Space Sciences, Faculty of Science,
  Ankara University, TR-06100 Tandogan, Ankara, Turkey}
\email{selam@science.ankara.edu.tr}

\begin{abstract}
We analyze new, high quality multicolor light curves of four overcontact
binaries: \object{AK Her}, \object{HI Dra}, \object{V1128 Tau} and
\object{V2612 Oph}, and determine their orbital and physical parameters
using the modeling program of G. Djurasevic and recently published results of
radial velocity studies. The achieved precision in absolute masses is between 10
and 20\%, and in absolute radii between 5 and 10\%. All four systems are W~UMa
type binaries with bright or dark spots indicative of mass and energy transfer
or surface activity. We estimate the distances and the ages of the systems
using the luminosities computed through our analysis, and perform an O-C study
for \object{V1128 Tau}, which reveals a complex period variation that can be
interpreted in terms of mass loss/exchange and either the presence of the third
body, or the magnetic activity on one of the components. We conclude that
further observations of these systems are needed to deepen our understanding of
their nature and variability.
\end{abstract}


\section{Introduction}

Overcontact binaries are close binary star systems in which one of the
components has become engulfed in the expanding envelope of its evolving
companion. The two stars are virtually indistinguishable and often appear to be
of same effective temperatures, since they share a common envelope and are
typically located at a separation of the same order of magnitude as the stellar
radii. However, the components of overcontact systems are stars of different
masses and at different evolutionary stages, with the added caveat that their
evolution has been affected by the proximity of their companion. In fact, it is
not uncommon to find overcontact systems in which the originally less massive
component is now observed to be the more massive one thanks to accumulating the
material lost during its companion`s evolution-driven expansion. This
phenomenon, often referred to as the mass ratio reversal, can in principle
happen more than once in the same system. Yet it's only one of the many
astrophysically interesting processes that can be inferred from the study
of overcontact systems, like the exchange of energy and angular momentum, mass
loss, interaction of stellar winds, magnetic activity and so forth. On the other
hand, the small separation makes it more probable that an overcontact system
will also be found to be eclipsing, which allows precise determination of
orbital and stellar parameters from photometric light curves through the
well-established methods for modeling of binary stars. Considering all this, it
is clear why the case studies of bright, eclipsing overcontact binaries remain
attractive even in the era of space telescopes and massive surveys.

In this paper we present the analysis of new, high quality CCD light curves,
based on the results of up to date radial velocity studies of four overcontact
eclipsing binaries: \object{AK Her}, \object{HI Dra}, \object{V1128 Tau} and
\object{V2612 Oph}. Our findings suggest that \object{AK Her} and
\object{V2612 Oph} belong to the A subtype of W~UMa binaries, while
\object{HI Dra} and \object{V1128 Tau} are of W-subtype. In all four solutions,
we use dark or bright spots on one or both components to explain the asymmetries
of the light curves. These inhomogeneities in surface brightness can be
explained as either arising from photospheric activity, or resulting from mass
and energy transfer.

In addition to constraining the absolute masses and radii of the components of
these four systems, we update their ephemerides using newly measured times
of minimum light combined with all the previous measurements found in
literature and calculate the distances based on the computed luminosities
resulting from our models. We then estimate the age of each system by
employing a novel method developed specifically for W UMa type binaries
\citep{yildiz}. The age calculation is based on the estimated initial mass of
the less massive secondary star, which can in turn be infered from the departure
of its size and luminosity from values expected in a main sequence star of its
mass (see the cited work and the references within for a detailed
explanation).

Finally, we perform a period study for \object{V1128 Tau} based on the O-C
diagram constructed with all the available times of minimum light. We find the
period variation can be fitted well with a superposition of a quadratic and a
cyclic function, indicative of both mass transfer and either a third body in the
system or magnetic activity on one of the components.

\begin{figure*}
\includegraphics{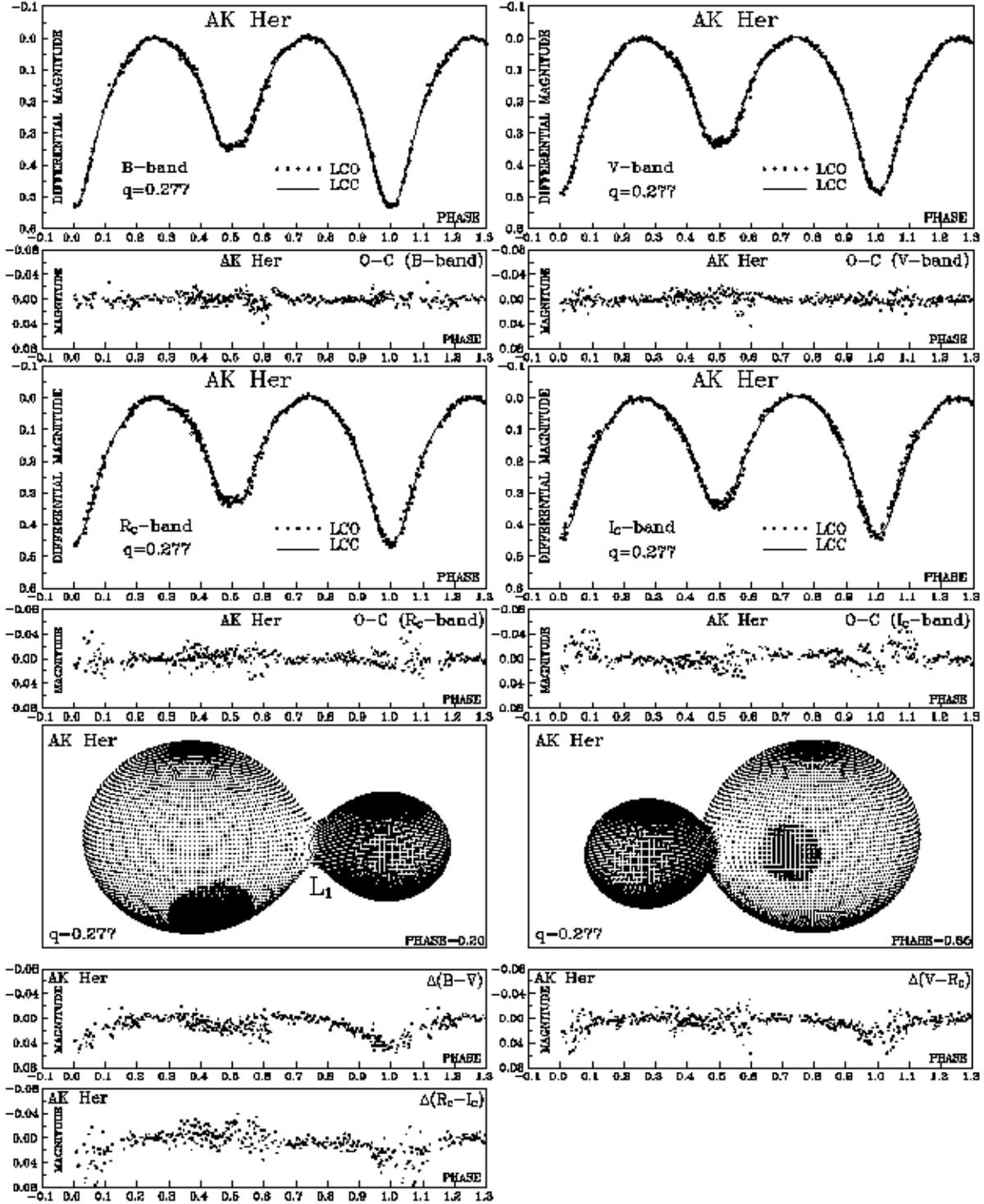}
\caption{Observed (LCO) and synthetic (LCC) light curves and the final O-C
	residuals of {\rm AK Her}, with $\Delta(B-V)$, $\Delta(V-R_C)$ and
	$\Delta(R_C-I_C)$ color curves and the graphic representation of the model
	described in Section~\ref{sAKHer} at the orbital phases 0.20 and 0.85.}
\label{fAKHer}
\end{figure*}

\begin{figure*}
\includegraphics{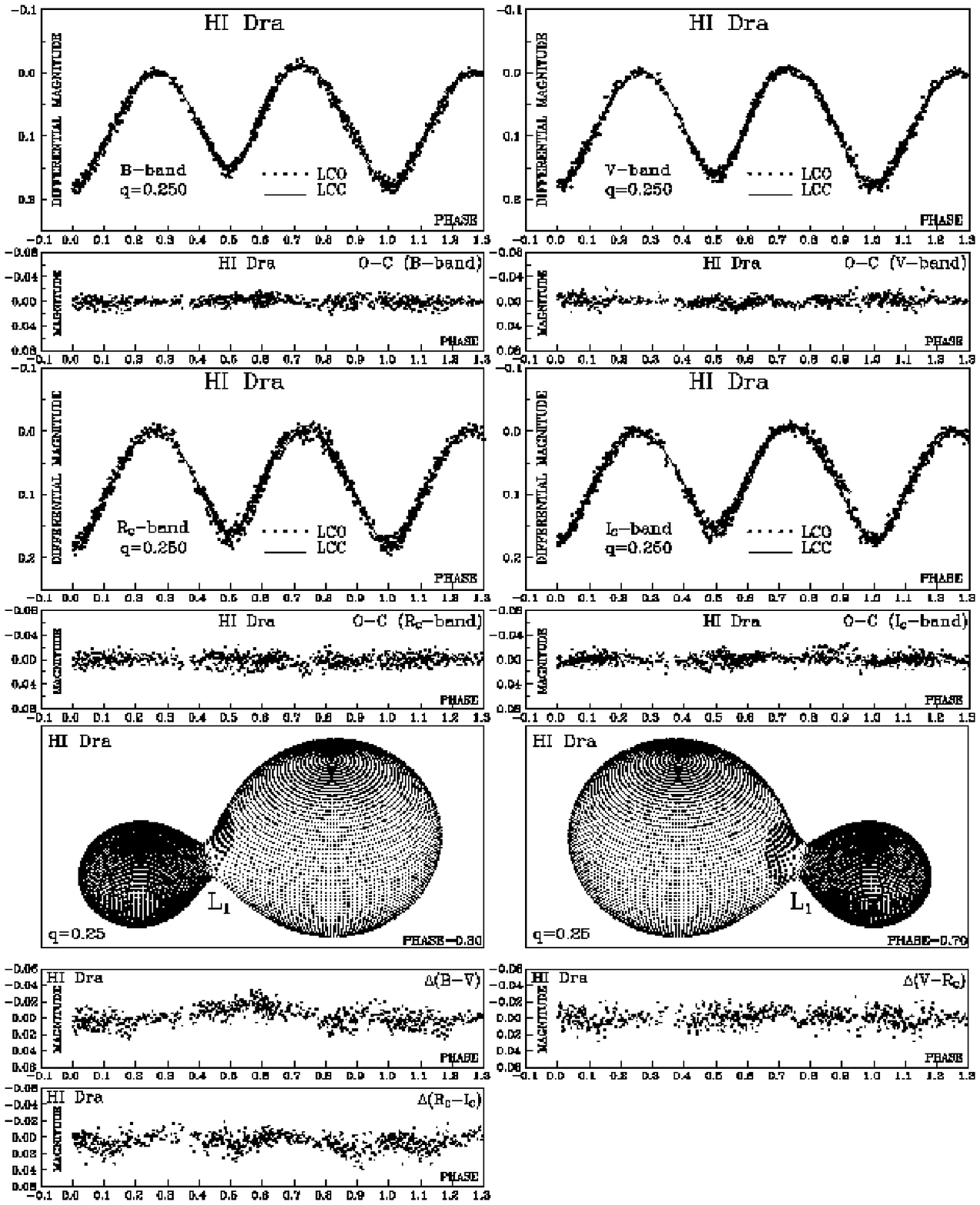}
\caption{Observed (LCO) and synthetic (LCC) light curves and the final O-C
	residuals of {\rm HI Dra}, with $\Delta(B-V)$, $\Delta(V-R_C)$ and
	$\Delta(R_C-I_C)$ color curves and the graphic representation of the model
	described in Section~\ref{sHIDra} at orbital phases 0.30 and 0.70.}
\label{fHIDra}
\end{figure*}

\begin{figure*}
\includegraphics{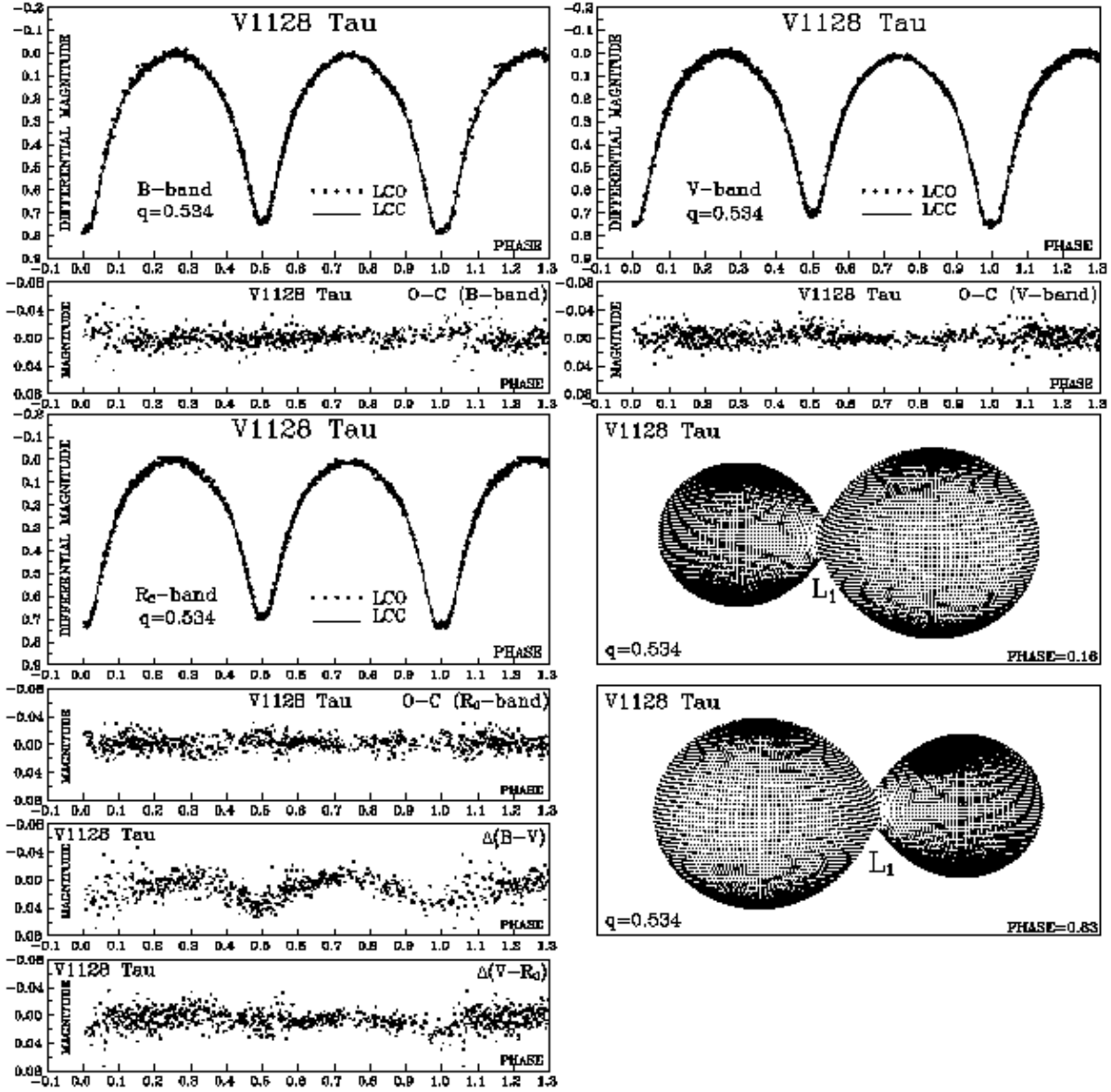}
\caption{Observed (LCO) and synthetic (LCC) light curves and the final O-C
	residuals of {\rm V1128 Tau}, with $\Delta(B-V)$ and $\Delta(V-R_C)$ color
	curves and the graphic representation of the model described in
	Section~\ref{sV1128Tau} at orbital phases 0.16 and 0.83.}
\label{fV1128Tau}
\end{figure*}

\begin{figure*}
\includegraphics{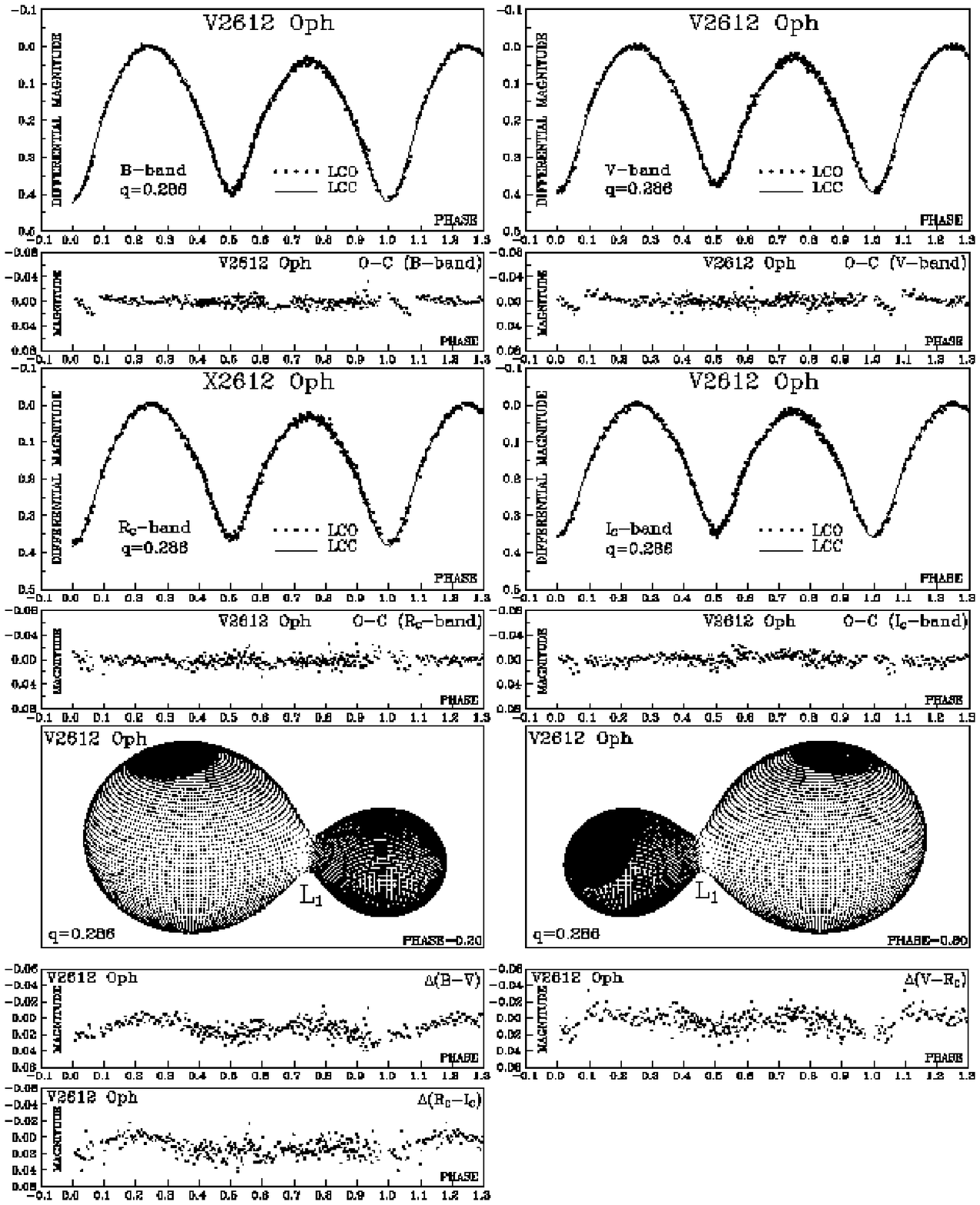}
\caption{Observed (LCO) and synthetic (LCC) light curves and the final O-C
	residuals of {\rm V2612 Oph}, with $\Delta(B-V)$, $\Delta(V-R_C)$ and
	$\Delta(R_C-I_C)$ color curves and the graphic representation of the model
	described in Section~\ref{sV2612Oph} at orbital phases 0.20 and 0.80.}
\label{fV2612Oph}
\end{figure*}

In what follows, we describe the technical details of the observations
(Section~\ref{observations}) and the light curve analysis
(Section~\ref{analysis}), and then each system is discussed in turn. A summary
of our results can be inspected at a glance from
Figures~\ref{fAKHer},~\ref{fHIDra},~\ref{fV1128Tau} and~\ref{fV2612Oph}, as
well as from the concluding remarks given in Section~\ref{resume}.

\section{Observations}\label{observations}

We obtained CCD photometric observations of our objects with the Apogee
ALTA U47 CCD camera attached to the 40~cm Schmidt-Cassegrain telescope at the
Ankara University Kreiken Observatory (AUKR), using the wideband BVRI filters of
Johnson-Cousins system.

The bias subtraction, dark and flat corrections were applied to the object
images by the IRAF\footnote{IRAF is distributed by the National Optical
Astronomy Observatories, which are operated by the Association of Universities
for Research in Astronomy, Inc., under cooperative agreement with the National
Science Foundation.} task CCDPROC. We then performed aparture photometry with
the relevant tasks in the IRAF/APPHOT package on each individual calibrated
frames. 

The magnitudes and their errors in each band were computed in the sense of
variable minus comparison (V-C). We determined the nightly extinction
coefficients for the magnitudes from the comparison stars. 

The light curves were phased using the light elements calculated from newly
observed and archival times of minima. The light elements, comparison stars,
and the uncertainties of the observations are listed in Table~\ref{tab-log},
and the times of minimum light newly derived from our observations
are given in Table~\ref{tab-min}. A preview of the data is given in
Table~\ref{tab-data}, and the full light curves are available as a
machine-readable table in the online version of the journal.

\section{Light curve analysis}\label{analysis}

Light curve analysis was done with the version of the program by
\citet{djur92a} generalized for the case of overcontact configurations
\citep{djur98}. The underlying model is based on the Roche geometry, and the
orbital and stellar parameters are estimated by solving the inverse problem with
a modified \citet{marquardt} algorithm. More details about the model and
parameter determination can be found in \citet{djur92b}.

Our results are summarized in Tables~\ref{tabAKHer}, \ref{tabHIDra},
\ref{tabV1128Tau}, and \ref{tabV2612Oph}. In the interest of limiting the
parameter space and better constraining the solution, certain parameters were
fixed to values obtained through independent analysis of radial velocity curves
or to theoretical values based on plausible assumptions about the components.
Below is a list with all the model parameters (with adopted values where
applicable) and explanations of used notation:

\renewcommand{\labelitemi}{$\circ$}
\begin{footnotesize}
\begin{itemize}
\itemsep0pt
\item Point count --- the total number of the observations spanning all the
  passbands.
\item $\rm \sigma$ --- the standard deviation of the residuals.
\item $\rm q=m_2/m_1$ --- the mass ratio of the components. The index $1$
  always marks the more massive component, so that the mass ratio is always
  less than one. This parameter was kept constant, with the value adopted from
  recent radial velocity studies: $q=0.277\pm0.024$ for AK Her, from
  \citet{AKHerRV}; $q=0.250\pm0.005$ for HI Dra, from \citet{HIDraRV};
  $q=0.534\pm0.006$ for V1128 Tau, from \citet{V1128TauRV}; and
  $q=0.286\pm0.003$ for V2162 Oph, from \citet{V2612OphRV}.
\item i --- the orbital inclination (in degrees).
\item $\rm a_{orb}$ --- the orbital semi-major axis in units of solar radius
  (where the value of $\rm a \cdot sin(i)$ adopted from the radial velocity
  studies mentioned above).
\item d --- the distance to the object in parsecs.
\item $\rm \ell_3/(\ell_1+\ell_2+\ell_3)$ --- the contribution of uneclipsed
  (third) light to the total light of the system at the phase of the light-curve
  maximum (omitted when zero).
\item $\rm f_{over}$ --- the degree of overcontact (in percents), defined as
  $f_{over}=100 \frac{\Omega - \Omega_{in}}{\Omega_{out} - \Omega_{in}}$,
  where $\rm \Omega,\ \Omega_{in}\ and\  \Omega_{out}$ are the dimensionless
  surface potentials of the common photosphere and the inner and outer
  critical surfaces, respectively (omitted in detached systems).
\item $\rm \Omega_{in}\ and\ \Omega_{out}$ --- the dimensionless values of
  Roche potential at the inner and outer critical surfaces that contain the
  equilibrium points $\rm L_1$ and $\rm L_2$, respectively (omitted in detached
  systems).
\item A, $\beta$ --- the albedo and the gravity-darkening exponent of the
  component. These parameters were fixed to their theoretical values
  \citep[see][]{lucy,vonz}, according to the the temperature inferred from the
  spectral classification for the primary, and a preliminary estimate of the
  temperature of the secondary.
\item $f=\omega/ \omega_K$ --- the ratio of the rotation rate of the component
  to the Keplerian orbital rate. This parameter was fixed to the value of
  $f=1$ for all systems, along the assumption that the rotation of the
  components is synchronous with the orbital revolution since the tidal
  effects are expected to lead to the synchronization of the rotational and
  orbital periods in such close systems.
\item $\rm T_{eff}$ --- the effective temperature of the component (in
  Kelvins), corresponding to the average of local temperatures weighted by the
  areas of elementary surfaces. It is usually estimated from spectral
  classification for one component (according to the revised theoretical $\rm
  T_{eff}$--spectral-type calibration by \citealt{mart05}), and adjusted as a
  parameter of the model for the other.
\item F --- the filling factor of the component, defined as the ratio between
  the stellar polar radius and the polar radius of the critical Roche surface.
\item $\rm \Omega$ --- the dimensionless surface potential of the component.
\item $\rm L/(L_1+L_2)$ --- the contribution of the component to the total
  luminosity of the system.
\item R --- the polar radius of the component in units of separation.
\item $\cal M, \ \cal R$ --- the mass and the mean radius of the component in
  solar units.
\item $\rm \log g$ --- the logarithm (to base ten) of the effective gravity of
  the component in CGS units.
\item $\rm M_{bol}$ --- the absolute bolometric magnitude of the component.
\item $\rm T_{spot}/T$ --- the ratio between the temperature of the spot and the
  local temperature of the star.
\item $\rm \theta,\ \lambda,\ \varphi$ --- the angular radius, longitude, and
  latitude of the spot (in arc degrees). The longitude ($\rm \lambda$) is
  measured in the orbital plane, from the $+x$ axis (the line connecting the
  centers of the components), clockwise as viewed from the "north" ($+z$)
  pole, in the range from $0\degr$ to $360\degr$. The latitude ($\varphi$) is
  measured from $0\degr$ at the the orbital plane to $90\degr$ towards the
  "north" ($+z$) pole and $-90\degr$ towards the "south" ($-z$) pole.
\end{itemize}
\end{footnotesize}

Limb darkening is calculated according to the nonlinear approximation of
\citet{claret00}, with the coefficients for the appropriate passbands
interpolated from their tables based on the values of ${\rm T_{eff}}$ and
$\rm \log~g$ in each iteration.

The distances are computed based on the apparent magnitudes taken from the
SIMBAD database\footnote{http://simbad.u-strasbg.fr/simbad/}, published by
\citet{vanlee07} for all stars except \object{V2612 Oph}, for which the SIMBAD
database has no published source; and from the computed absolute magnitudes,
with corrections for the interstellar extinction. The interstellar extinction
values ($A_v$) in the V passband are calculated using the reddening values
estimated from the infrared dust emission maps of \citet{schlegel} and by
assuming an extinction to reddening ratio of $A_v/E(B-V)$ of 3.1. Note that
\citet{schlegel} refer to the total absorption. The bolometric corrections are
taken from the tables published by \citet{flower}, according to the
computed effective temperatures of the components for each system.

We use the methods prescribed by \citet{yildiz} to estimate the inital masses
and ages of the systems, and classify them as either the A or W subtypes of W
UMa binaries according to the criteria introduced by \citet{binnen}.

The uncertainties reported in Tables \ref{tabAKHer}, \ref{tabHIDra},
\ref{tabV1128Tau}, \ref{tabV2612Oph} and throughout the text are derived from
the formal fitting errors.

In the following sections we present the results of our work for each system in
turn.

\section{AK Her}\label{sAKHer}

\object{AK Her} (HD 155937, BD+16 3130, HIP 84293, SAO 102688, GSC 1536-1738)
is an overcontact eclipsing binary with the orbital period of $P=0.421522$
days. It's the brighter component of the visual binary ADS 10408. The visual
companion is a physical member of the system, separated by 4.7 arcseconds and
3.5 mag fainter than \object{AK Her} at maximum light. It has been associated
with a weak X-ray source \citep{cruddace, mcgale}, which is an indication of
coronal activity. Numerous authors published photometric light curves and times
of minimum light. The variability of the orbital period was first noted by
\citet{woodward41}. \citet{schmidt} did a through analysis of the variations
in the light curves during the years prior to their study. They pointed out
the changing levels of light curve minima and maxima, as well as the variability
in spectral types and color indices, and ruled out the distant visual companion
as the cause of period variation. A detailed study of the period changes was
performed by \citet{li}, who found a long-term decrease and three rapid
components of the variation. They interpret the slow variability in terms of
mass exhange, and the rapid variations as pulsations of the common envelope.
More recently, \citet{samadi} analyzed multicolor light curves of
\object{AK Her} together with the radial velocity data from \citet{sanford}
and calculated the absolute parameters. Based on the O-C analysis, they measure
the amplitudes and frequencies of the three componenst of orbital period
variation, and interpret them in terms of the presence of the third body and
magnetic activity.

A spectroscopic study of \object{AK Her} was done by \citet{AKHerRV}. The
authors classified the brighter star as F4~V and measured the spectroscopic mass
ratio of $q=0.277$, which is the value used in this work. Although they did not
detect the signature of additional components in the spectra, they too argue,
based on the observed light-time effect, that there must be an unseen companion
to the system much closer than the visual pair. However, it has not yet been
directly observed.

The present analysis of BVR$\rm _C$I$\rm _C$ light curves results in a model
of \object{AK Her} in which the more massive, larger and hotter (primary) star
is the one eclipsed in the deeper minimum (Figure~\ref{fAKHer}). This makes
\object{AK Her} a member of the A subtype of W UMa binaries. From the summary of
results given in Table~\ref{tabAKHer} it can be seen that the mass and the
radius of the primary are near the expected values for a main sequence star of
spectral class F4~V, adopted from \citet{AKHerRV}, while the secondary is too
large and overluminous for a main sequence star of its mass. This is a common
observational characteristic of overcontact systems. We find the degree of
overcontact is 33.2\%, and place two dark spots that likely arise from surface
activity on the primary star to account for the light curve assymetries. There
is a significant contribution of uneclipsed light, probably from the unresolved
third body.

These results are in reasonable agreement with the findings of previous
studies. In comparison with the most recent one, done by \citet{rovithis}, we
report a larger inclination ($i_{R-L}=76\degr$ compared to $82\degr$ in the
present work), a slightly higher temperature of the secondary ($T_{R-L}$=5900
K, compared to 6200 K in this study), and different polar radii of the
components, but within our estimated uncertainties.

The distance of the system, according to the absolute magnitude from
Table~\ref{tabAKHer}, is 103 $\pm$ 3 pc. Based on the current mass and
luminosity of the secondary component and the methods described by
\citet{yildiz}, we estimate that the age of \object{AK Her} is 4.58 $\pm$ 2.51
Gyr.

\section{HI Dra}\label{sHIDra}

\object{HI Dra} (HD 171848, BD+58 1824, NSVS 3037033, GSC 03917-02301, ADS
11465 A) is a bright contact binary  discovered in the Hipparcos observations
and initially classified as an RR~Lyr variable. \citet{gomez} published the
light curve and suggested that HI Dra may be a $\beta$~Lyr or an ellipsoidal
variable, pointing that the difference of 0.02 mag between light curve levels
prior and after the primary minimum is indicative of binarity. \citet{selam}
re-analyzed the Hipparcos photometric data under the assumption that the
object is a binary system, and determined a photometric mass ratio of 0.15 and
an inclination angle of 52.5 degrees.

The first spectroscopic study of \object{HI Dra}, done by \citet{HIDraRV},
confirmed the binary nature of the system. The authors calculated the
spectroscopic mass ratio of $q=0.25$, which is the value used in the present
work, and determined the spectral class as F0$-$F2~V.

The results of our analysis of the new BVR$\rm _C$I$\rm _C$ light curves of
\object{HI Dra} are summarized in Table~\ref{tabHIDra} and illustrated in
Figure~\ref{fHIDra}. Due to the very low inclination ($i=54^{\degr}$), the
light curve is of sinusoidal shape reminiscent of ellipsoidal variables, with
minima of nearly equal depth. The slight asymmetry of the light curves is
accounted for by adding a bright spot to the primary star in the neck region
of the system. Since it is the less massive secondary that is eclipsed in the
deeper minimum, this is a W-type W UMa binary with the degree of overcontact
of 23\%. Within the estimated uncertainty, the primary has a mass that could
be expected from a class F0-F2 main sequence star, but a larger radius and
luminosity. We find a faint third light in B and I$\rm _C$ bands. Estimated
distance of \object{HI Dra}, based on the absolute parameters resulting from our
analysis, is 304 $\pm$ 27 pc, and its age, 2.01 $\pm$ 1.18 Gyr.

One of the intermediate results of the age calculation is the initial mass of
the secondary component, which in this case 2.14 M$_\odot$. According to
\citet{yildiz}, this indicates that the system should have evolved into an
A-subtype W UMa binary. However, when we made trial models with the A-subtype
configuration, could not obtain a fit of satisfying quality, and the best
solutions in such a case demanded greatly increased temperature of the
secondary (in excess of 1000 K above the temperature of the primary), which
isn't characteristic of W~UMa stars. Therefore, we remain with the more likely
classification of \object{HI Dra} as a W-subtype system.

\section{V1128 Tau}\label{sV1128Tau}

\object{V1128 Tau} is a W~UMa type variable with a visual companion
(BD+12~511B) separated by 14 arcsec. Its light curves were studied by
\citet{tas}, \citet{hawkins} and \citet{zhang} with relatively homogenous
conclusions. Using newly measured times of minimum light and all the available
data from the literature, \citet{liu} reported a long-term variation of the
orbital period, which could be explained by Applegate mechanism on one of the
components. \citet{V1128TauRV} performed a radial velocity study and found a
mass ratio of $q=0.534$, which is used in the present work, and estimated a
spectral type of F8~V.

Our analysis of the new BVR$\rm _C$ light curves of \object{V1128 Tau} is
summarized in Table~\ref{tabV1128Tau} and illustrated in Figure~\ref{fV1128Tau}.
The light curves exhibit continual changes, minima of almost equal depths, and a
slight asymmetry between maxima that we account for by putting a dark spot on
the secondary component. The mass and radius of the primary are reasonably close
to those expected from a main sequence F8 star. Since it is the more massive
star that is eclipsed in the deeper minimum, \object{V1128 Tau} is a W-type
W~UMa system. We find that the degree of overcontact is 13.4\%, the distance is
139 $\pm$ 3 pc, and the age of \object{V1128 Tau} is 5.04 $\pm$ 1.86 Gyr.

Our results are in good agreement with the most recent photometric study of
\object{V1128 Tau} done by \citet{zhang}, with the exception of the presence of
the dark spot; namely, they presented a spotless solution, whereas we were not
able to obtain a satisfying fit to the observations without the inclusion of a
spot.

\subsection{O-C analysis}

\begin{figure*}
\includegraphics{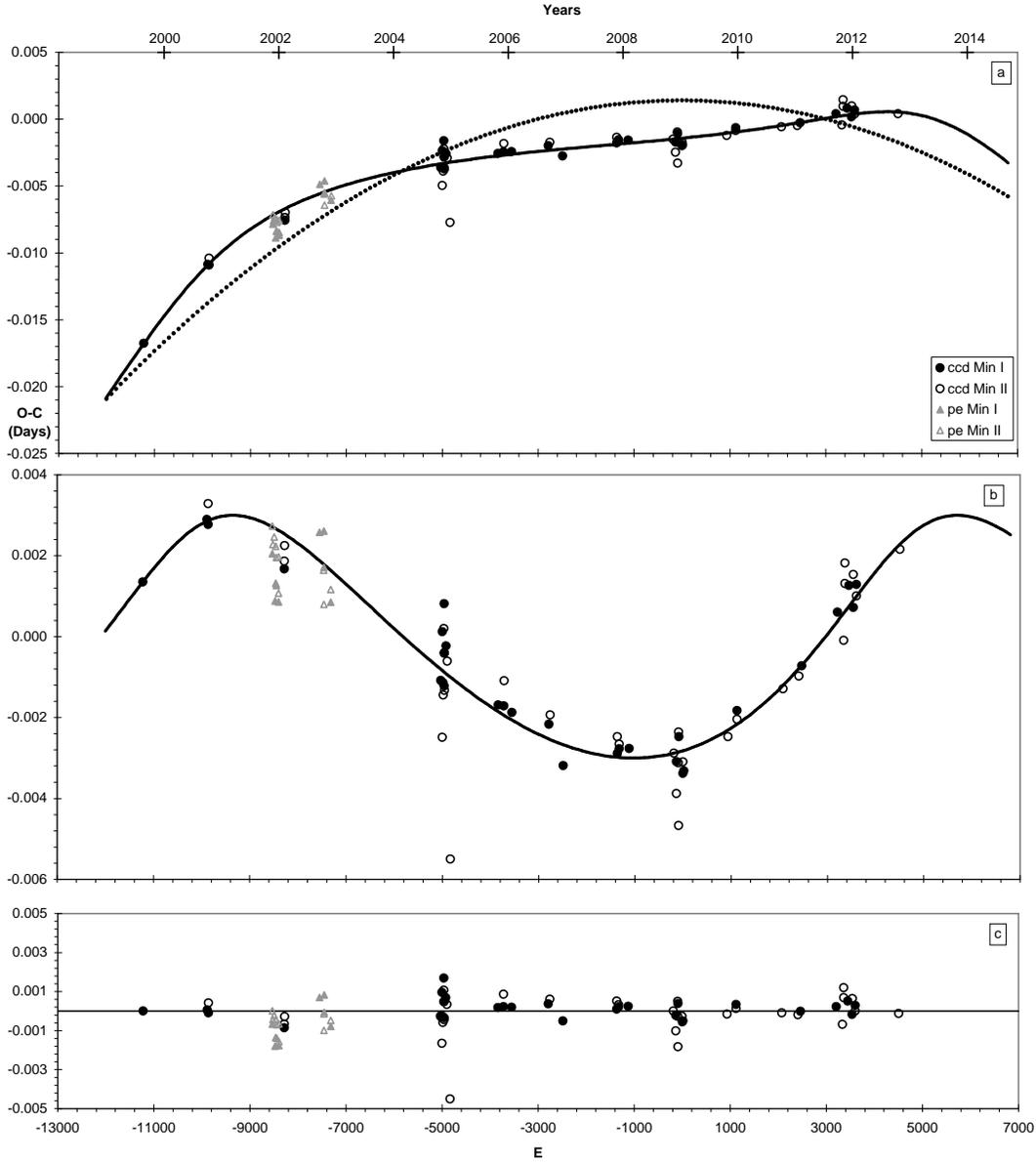}
\caption{(a) The O-C diagram for V1128 Tau with the photoelectric (gray
	triangles) and CCD (black circles) times of minima. The dotted and solid
	curves show the quadratic and overall (quadratic plus cyclic) parts of the
	fitted function. (b) The residuals from the quadratic fit;
	the solid curve represents the cyclic part of the fitted function.
	(c) The O-C residuals after the subtraction of the fitted function.}
\label{fV1128TauOC}
\end{figure*}

We performed an O-C analysis of V1128 Tau based on the photoelectric and CCD
times of minimum light that we collected from the literature and observed
ourselves. The O-C diagram (see Fig.~\ref{fV1128TauOC}),
displays a cyclic character superimposed on a quadratic variation. The quadratic
variation can be explained by the mass exchange/mass loss in the system, while
cyclic variation may be the result of the presence of a gravitationally bound
third body or magnetic activity on one of the components
\citep[see e.g.][]{selamandalbayrak07}.

The orbital period decrease of $dP/dt=-3.46\times10^{-8}$ day\,
yr$^{-1}$ is indicative of mass transfer from the primary to the secondary
component at a rate of $\sim-1.13\times10^{-7}M_{\odot}y^{-1}$, unless mass
is being lost from the system through some other mechanism, e.g. stellar winds.

Considering a hypothetical third body gravitationally bound to the close binary,
we determined the parameters of the light-time orbit as
$T_{0}$, $P_{orb}$, $\frac{dP}{dE}$, $P_{12}$, $T^{'}$, $a^{'}_{12}\sin i^{'}$,
$e^{'}$, $\omega ^{'}$, using the OC2LTE30 code \citep{aketal04}, which is based
on the formulation by \citet{irwin52}.

The magnetic activity parameters of the secondary component can be calculated
using the formalism of \citet{applegate92} as an alternative explanation for
the observed cyclic period variation. In this context, the period and amplitude
of the cyclic period variation are found to be 12.54\,years and 0.105\ s\
cyc$^{-1}$ respectively, while the angular momentum transfer and the energy
required for the angular momentum transfer are computed to be
$-8.34\times10^46$ g\,cm$^{2}$\,s$^{-1}$ and $6.46\times10^{40}$ erg. The
resulting variations in the luminosity and brightness of the secondary are then
$5.13\times10^{32}$ erg\,s$^{-1}$ and $0.{^m}01$ respectively. All the
parameters of the suggested quadratic and cyclic period variations are
summarized in Table~\ref{tabOC}.

The O-C diagram for V1128 Tau, based on all observations made so far, covers one
incomplete cycle of the predicted variation. We thus refrain from drawing any
definite coclusions about the mechanism for the suggested cyclic variation.
Further observations are needed to understand its nature and origin.

\section{V2612 Oph}\label{sV2612Oph}

\object{V2612 Oph} (HD 170451, BD+06 3809, NSV 10892, GSC 445-1993, NGC
6633-147) is bright (V$_{max}$ = 6$^{m}$.20) eclipsing binary that was thought
to be a member of the galactic cluster \object{NGC 6633} \citep{hiltner}. Its
light curves were studied by several groups \citep{koppelman, hidas, yang, deb}.
\citet{V2612OphRV} did a spectroscopic analysis and determined the
spectral class of the system as F7~V and its mass ratio as $q=0.286$, which is
the value used in the present work. They dispute the cluster membership of
\object{V2612 Oph} on the basis of the detection of a much fainter
(V = 12$^{m}$.8) W-UMa type cluster member, whereas the brightness of
\object{V2612 Oph}, when calculated using the distance modulus of
\object{NGC 6633} given by \citet{kharchenko}, would have to be
V$_{max}$ = 11$^{m}$.26 if it was a member of the cluster.

The new BVR$\rm _C$I$\rm _C$ light curves of \object{V2612 Oph}, along with the
best fitting model, are shown in Figure~\ref{fV2612Oph}, and the summary of our
analysis is given in Table~\ref{tabV2612Oph}. There's a significant difference
in the height of adjacent maxima, as well as a noticable asymmetry. These
features are modeled by placing a dark spot in the polar region of each
component. According to \citet{solanki}, the presence of spots at high latitudes
can be explained by the dynamo mechanism for rapid rotators, and short-period,
close binaries can be considered as such. The absolute parameters of the primary
are in accordance with its spectral classification as a main sequence star.
As the deeper minimum corresponds to the eclipse of the more massive primary
star, this is an A-subtype W~UMa binary. The degree of overcontact is 22\%,
the distance is 173 $\pm$ 5 pc, and the age of \object{V2612 Oph} is 5.11 $\pm$
1.79 Gyr. The distance and age of \object{V2612 Oph} indicate that it should not
be a member of \object{NGC 6633} cluster, which has a distance of 376~pc and age
of 425~Myr (according to WEBDA\footnote{http://www.univie.ac.at/webda/}).

These findings are in good agreement with the results of previous studies,
with one exception: \citet{deb} classifies \object{V2612 Oph} as a W-subtype
W~UMa system. Such classification is not supported by our model, which cannot
fit the observations well if a W-subtype configuration is attempted.

\section{Resume}\label{resume}

We performed the numerical analysis of four W~UMa type overcontact
binaries: \object{AK Her}, \object{HI Dra}, \object{V1128 Tau}, and
\object{V2612 Oph} and determined their absolute physical and orbital
parameters. Our results are based on new photometric observations and
recent radial velocity studies. The analysis was done using a sophisticated
model of a binary system that provided excellent fits to the observations. In
addition to updating the ephemeris and the absolute parameters, we calculated
the distance and estimated the age of each object, and did an O-C analysis for
\object{V1128 Tau} that revealed a complex period variation consistent with
mass transfer within the system and either the presence of the third body or
magnetic activity on one of the components. Unfortunately, at the time of this
writing, the available times of minimum light for \object{HI Dra} and
\object{V2612 Oph} do not allow for the construction of an O-C diagram. This
could be achieved in the future with further observations, which would help to
better understand the nature and variability of these systems, and widen the
knowledge about overcontact binary stars in general.

\acknowledgments

This research was funded in part by the Ministry of Education, Science and
Technological Development of Republic of Serbia through the project "Stellar
Physics'' (No. 176004). The authors acknowledge the use of the
\url[http://simbad.u-strasbg.fr/simbad/]{Simbad database}, operated at the
CDS, Strasbourg, France, \url[http://adsabs.harvard.edu/]{NASA's
Astrophysics Data System Bibliographic Services} and
\url[http://www.univie.ac.at/webda/] {WEBDA database of stellar clusters in the
Galaxy and the Magellanic Clouds}.

{\it Facilities:} \facility{AUKR}.

\clearpage

\begin{table*}
\caption{Summary of observational aspects.
\label{tab-log}}
\begin{footnotesize}
\begin{tabular}{llccccll}
\tableline\tableline
Program star	& Comparison star 	& \multicolumn{4}{c}{Light curve $\sigma$}
	& Epoch [HJD]		& Period [days] \\
 				& 					& B 	& V 	& R 	& I
	& 					& 				\\
\tableline
AK Her 			& {PPM 133147}		& 0.152 & 0.144 & 0.144 & 0.133
	& 2453176.393(9)	& 0.421523(2) 	\\
HI Dra 			& {GSC 3917-1556} 	& 0.057 & 0.058 & 0.058 & 0.056
	& 2456534.419(2)  	& 0.597423(2) 	\\
V1128 Tau 		& {GSC 664-387} 	& 0.243 & 0.226 & 0.221 &  ---
	& 2454842.3470(1) 	& 0.305371(2) 	\\
V2612 Oph 		& {GSC 445-1293} 	& 0.126 & 0.119 & 0.115 & 0.112
	& 2453846.916(1)  	& 0.375307(3) 	\\
\tableline
\end{tabular}
\end{footnotesize}
\end{table*}

\begin{table}
\caption{New times of minimum light derived from our observations.
\label{tab-min}}
\begin{footnotesize}
\begin{tabular}{lll}
\tableline\tableline
System 		& Time of minimum [HJD] & Type 		\\
\tableline
AK Her 		& 2456512.3234(3)		& Primary	\\
 			& 2456516.3284(1)		& Secondary \\
HI Dra 		& 2456526.3530(1) 		& Secondary \\
 			& 2456534.4190(2) 		& Primary 	\\
 			& 2456372.5190(1) 		& Primary 	\\
V1128 Tau	& 2454785.3952(1)		& Secondary \\
			& 2454813.4894(1)	 	& Secondary \\
			& 2454842.1949(1)	 	& Secondary \\
			& 2454842.3480(1) 		& Primary 	\\
V2612 Oph 	& 2456517.4112(1) 		& Secondary \\
 			& 2456519.2882(1) 		& Secondary \\
			& 2456530.3591(2)       & Primary 	\\
\tableline
\end{tabular}
\end{footnotesize}
\end{table}

\begin{table}
\caption{BVRI Photometry of AK Her, HI Dra, V1128 Tau, V2612 Oph}
\label{tab-data}
\begin{footnotesize}
\begin{tabular}{llc}
\tableline\tableline
\multicolumn{3}{c}{AK Her: \textit{B} band}                      \\
\tableline
HJD (d)			&   Phase   &   Diff. Magnitude (mag)	\\
2456510.2891	& 	0.1714	&	-1.006                  \\
2456510.2905	&	0.1747	&	-1.013                  \\
2456510.2919	& 	0.1781	&	-1.017                  \\
2456510.2933	& 	0.1814	&	-1.016                  \\
2456510.2947	& 	0.1848	&	-1.019                  \\
\tableline
\end{tabular}
\tablecomments{This table is available in its entirety in machine-readable and
Virtual Observatory (VO) forms in the online journal. A portion is shown here
for guidance regarding its form and content.}
\end{footnotesize}
\end{table}

\begin{table}
\caption{Summary of modeling results for AK Her. \label{tabAKHer}}
\begin{scriptsize}
\begin{tabular}{lll}
\tableline\tableline
Properties of the fit & &\\
\tableline
Point count & 1407  & \\
$\rm \sigma$ & 0.0097 & \\
\tableline\tableline
System parameters & & \\
\tableline
q & 0.277 $\pm$ 0.024 & \\
$\rm i\ [\degr]$ & 81.7  $\pm$ 0.2 & \\
$\rm a_{orb}\ [R_{\odot}]$ & 2.73 $\pm$ 0.09 & \\
d [pc] & 103 $\pm$ 3 \\
$\rm f_{over} [\%]$ & 33.2 & \\
$\rm \Omega_{in}, \Omega_{out}$ & 2.4148, 2.2412 & \\
$\rm \ell_3/(\ell_1+\ell_2+\ell_3)$ & \multicolumn{2}{l}{0.102 $\pm$ 0.002 [B],
	0.130 $\pm$ 0.002 [V], 0.146 $\pm$ 0.002 [$\rm R_C$], 0.162 $\pm$ 0.003
	[$\rm I_C$]} \\
\tableline\tableline
Stellar Parameters & Primary & Secondary \\
\tableline
A & 0.5 & 0.5 \\
$\rm \beta$ & 0.08 & 0.08 \\
f & 1.0 & 1.0 \\
$\rm T_{eff}$ [K] & 6500 &  6180  $\pm$ 10 \\
F & 1.027 $\pm$ 0.001 & 1.027 $\pm$ 0.001 \\
$\Omega$ & 2.3572 & 2.3572 \\
$\rm L/(L_1+L_2)$ & 0.786 [B], 0.782 [V], 0.780 [$\rm R_C$], 0.778 [$\rm I_C$]
	& 0.214 [B], 0.218 [V], 0.220 [$\rm R_C$], 0.222 [$\rm I_C$] \\
R [D=1] & 0.475 & 0.269 \\
$\rm L\ [L_{\odot}]$ & 3.02 $\pm$ 0.02 & 0.82 $\pm$ 0.02 \\
$\cal M\ \rm [M_{\odot}]$ & 1.2 $\pm$ 0.2 & 0.34 $\pm$ 0.07 \\
$\cal R\ \rm [R_{\odot}]$ & 1.40 $\pm$ 0.05 & 0.80 $\pm$ 0.03 \\
$\rm log_{10}(g)$ & 4.23 $\pm$ 0.08 & 4.2 $\pm$ 0.2 \\
$\rm M_{bol}$ & 3.54 $\pm$ 0.07 & 4.96 $\pm$ 0.08 \\
\tableline\tableline
Spot parameters & Spot 1 (Primary) & Spot 2 (Primary) \\
\tableline
$\rm T_{spot}/T$ & 0.807 $\pm$ 0.004 & 0.810 $\pm$ 0.06 \\
$\rm \theta\ [\degr]$ & 22.1 $\pm$ 0.2 & 16.1 $\pm$ 0.2 \\
$\rm \lambda\ [\degr]$ & 54.6 $\pm$ 0.7 & 319 $\pm$ 2 \\
$\rm \varphi\ [\degr]$ & -37.9 $\pm$ 0.6 & -1 $\pm$ 2 \\
\tableline
\end{tabular}
\end{scriptsize}
\end{table}

\begin{table}
\caption{Summary of modeling results for HI Dra. \label{tabHIDra}}
\begin{scriptsize}
\begin{tabular}{lll}
\tableline\tableline
Properties of the fit & &\\
\tableline
Point count & 2491 & \\
$\rm \sigma$ & 0.0076 & \\
\tableline\tableline
System parameters & & \\
\tableline
q & 0.250 $\pm$ 0.005 & \\
$\rm i\ [\degr]$ & 54.74 $\pm$ 0.05 & \\
$\rm a_{orb}\ [R_{\odot}]$ & 3.8 $\pm$ 0.2 & \\
d [pc] & 304 $\pm$ 27 & \\
$\rm f_{over} [\%]$ & 23.0 & \\
$\rm \Omega_{in}, \Omega_{out}$ & 2.3529, 2.1952 & \\
$\rm \ell_3/(\ell_1+\ell_2+\ell_3)$ & \multicolumn{2}{l}{0.032 $\pm$ 0.004 [B],
	0.000 $\pm$ 0.004 [V], 0.000 $\pm$ 0.004 [$\rm R_C$], 0.044 $\pm$ 0.004
	[$\rm I_C$]} \\
\tableline\tableline
Stellar Parameters & Primary & Secondary \\
\tableline
A & 0.5 & 0.5 \\
$\rm \beta$ & 0.08 & 0.08 \\
f & 1.0 & 1.0 \\
$\rm T_{eff}$ [K] & 7000 & 6550 $\pm$ 20 \\
F & 1.017 $\pm$ 0.001 & 1.017 $\pm$ 0.001 \\
$\rm \Omega$ & 2.3167 & 2.3167 \\
$\rm L/(L_1+L_2)$ & 0.822 [B], 0.818 [V], 0.815 [$\rm R_C$], 0.813 [$\rm I_C$] &
	0.178 [B], 0.182 [V], 0.185 [$\rm R_C$], 0.187 [$\rm I_C$]\\
R [D=1] & 0.478 & 0.257 \\
$\rm L\ [L_{\odot}]$ & 7.87 $\pm$ 0.04 & 1.80 $\pm$ 0.05 \\
$\cal M\ \rm [M_{\odot}]$ & 1.7 $\pm$ 0.3 & 0.42 $\pm$ 0.07 \\
$\cal R\ \rm [R_{\odot}]$ & 1.97 $\pm$ 0.09 & 1.07 $\pm$ 0.05 \\
$\rm log_{10}(g)$ & 4.1 $\pm$ 0.1 & 4.0 $\pm$ 0.2 \\
$\rm M_{bol}$ & 2.5 $\pm$ 0.1 & 4.1 $\pm$ 0.2 \\
\tableline\tableline
Spot parameters & Spot 1 (Primary) & \\
\tableline
$\rm T_{spot}/T$ & 1.139 $\pm$ 0.002 \\
$\rm \theta\ [\degr]$ & 15.5 $\pm$ 0.2 \\
$\rm \lambda\ [\degr]$ & 5.0  $\pm$ 0.3 \\
$\rm \varphi\ [\degr]$ & 6.8 $\pm$ 0.4 \\
\tableline
\end{tabular}
\end{scriptsize}
\end{table}

\begin{table}
\caption{Summary of modeling results for V1128 Tau. \label{tabV1128Tau}}
\begin{scriptsize}
\begin{tabular}{lll}
\tableline\tableline
Properties of the fit & & \\
\tableline
Point count & 1865 & \\
$\rm \sigma$ & 0.0098 & \\
\tableline\tableline
System parameters & & \\
\tableline
q & 0.534 $\pm$ 0.006 & \\
$\rm i\ [\degr]$ & 86.0 $\pm$ 0.2 & \\
$\rm a_{orb}\ [R_{\odot}]$ & 2.27 $\pm$ 0.04 & \\
d [pc] & 139 $\pm$ 3 \\
$\rm f_{over} [\%]$ & 13.4 & \\
$\rm \Omega_{in}, \Omega_{out}$ & 2.9406, 2.6240 & \\
\tableline\tableline
Stellar Parameters & Primary & Secondary \\
\tableline
A & 0.5 & 0.5 \\
$\rm \beta$ & 0.08 & 0.08 \\
f & 1.0 & 1.0 \\
$\rm T_{eff}$ [K] & 6200 & 6400 $\pm$ 10 \\
F & 1.017 $\pm$ 0.001 & 1.017 $\pm$ 0.001 \\
$\rm \Omega$ & 2.8982 & 2.8982 \\
$\rm L/(L_1+L_2)$ & 0.603 [B], 0.606 [V], 0.608 [$\rm R_C$]
	& 0.397 [B], 0.394 [V], 0.392 [$\rm R_C$] \\
R [D=1] & 0.416 & 0.312 \\
$\rm L\ [L_{\odot}]$ & 1.294 $\pm$ 0.009 & 0.839 $\pm$ 0.008 \\
$\cal M\ \rm [M_{\odot}]$ & 1.10 $\pm$ 0.06 & 0.58 $\pm$ 0.04 \\
$\cal R\ \rm [R_{\odot}]$ & 1.01 $\pm$ 0.02 & 0.76 $\pm$ 0.02 \\
$\rm log_{10}(g)$ & 4.47 $\pm$ 0.04 & 4.44 $\pm$ 0.04 \\
$\rm M_{bol}$ & 4.46 $\pm$ 0.04 & 4.93 $\pm$ 0.04 \\
\tableline\tableline
Spot parameters & Spot 1 (Secondary) & \\
\tableline
$\rm T_{spot}/T$ & 0.869 $\pm$ 0.006 \\
$\rm \theta\ [\degr]$ & 24.8 $\pm$ 0.5 \\
$\rm \lambda\ [\degr]$ & 146 $\pm$ 3 \\
$\rm \varphi\ [\degr]$ & 52 $\pm$ 2 \\
\tableline
\end{tabular}
\end{scriptsize}
\end{table}

\begin{table}
\caption{Summary of modeling results for V2612 Oph. \label{tabV2612Oph}}
\begin{scriptsize}
\begin{tabular}{lll}
\tableline\tableline
Properties of the fit & &\\
\tableline
Point count & 1363 & \\
$\rm \sigma$ & 0.0074 & \\
\tableline\tableline
System parameters & & \\
\tableline
q & 0.286 $\pm$ 0.003 & \\
$\rm i\ [\degr]$ & 66.66  $\pm$ 0.05 & \\
$\rm a_{orb}\ [R_{\odot}]$ & 2.59 $\pm$ 0.07 & \\
d [pc] & 173 $\pm$ 5 & \\
$\rm f_{over}$ [\%] & 22.1 & \\
$\rm \Omega_{in},\ \Omega_{out}$ & 2.4351, 2.2562 & \\
$\rm \ell_3/(\ell_1+\ell_2+\ell_3)$ & \multicolumn{2}{l}{0.000 $\pm$ 0.002 [B],
	0.004 $\pm$ 0.002 [V], 0.000 $\pm$ 0.003 [$\rm R_C$], 0.025 $\pm$ 0.003
	[$\rm I_C$]} \\
\tableline\tableline
Stellar Parameters & Primary & Secondary \\
\tableline
A & 0.5 & 0.5 \\
$\rm \beta$ & 0.08 & 0.08 \\
f & 1.0 & 1.0 \\
$\rm T_{eff}$ [K] & 6250 & 6280 $\pm$ 10 \\
F & 1.018 $\pm$ 0.001 & 1.018 $\pm$ 0.001 \\
$\rm \Omega$ & 2.3956 & 2.3956 \\
$\rm L/(L_1+L_2)$ & 0.751 [B], 0.751 [V], 0.752 [$\rm R_C$], 0.752 [$\rm I_C$]
	& 0.249 [B], 0.249 [V], 0.248 [$\rm R_C$], 0.248 [$\rm I_C$]\\
R [D=1] & 0.468  & 0.267  \\
$\rm L\ [L_{\odot}]$ & 2.23 $\pm$ 0.02 & 0.75 $\pm$ 0.01 \\
$\cal M\ \rm [M_{\odot}]$ & 1.3 $\pm$ 0.1 & 0.37 $\pm$ 0.04 \\
$\cal R\ \rm [R_{\odot}]$ & 1.30 $\pm$ 0.04 & 0.75 $\pm$ 0.01 \\
$\rm log_{10}(g)$ & 4.32 $\pm$ 0.06 & 4.25 $\pm$ 0.06 \\
$\rm M_{bol}$ & 3.87 $\pm$ 0.06 & 5.05 $\pm$ 0.06 \\
\tableline\tableline
Spot parameters & Spot 1 (Primary) & Spot 2 (Secondary) \\
\tableline
$\rm T_{spot}/T$ & 0.798 $\pm$ 0.003 & 0.807 $\pm$ 0.005  \\
$\rm \theta\ [\degr]$ & 28.3 $\pm$ 0.3 & 37.9 $\pm$ 0.2 \\
$\rm \lambda\ [\degr]$ & 168.7 $\pm$ 0.6 & 329.1 $\pm$ 0.5 \\
$\rm \varphi\ [\degr]$ & 77.9 $\pm$ 0.2 & 33 $\pm$ 1 \\
\tableline
\end{tabular}
\end{scriptsize}
\end{table}

\begin{table}
\caption{Parameters derived from the O-C analysis of V1128~Tau.
\label{tabOC}}
\begin{footnotesize}
\begin{tabular}{ll}
\tableline\tableline
Parameters & Values \\
\tableline
$\rm T_{0}$~[HJD]         			&$2454842.3470	\pm 0.0001$\\
$\rm P_{orb}$~[day]       			&$0.305371		\pm 0.0000001$\\
$\frac{dP}{dE}$~[day/cyc] 			&$-3.10			\pm 0.02\times10^{-10}$  \\
$\rm a^{'}_{12}\sin i^{'}$~[AU]     &$0.53			\pm 0.04$ \\
$\rm e^{'}$                   		&$0.3           \pm 0.2 $\\
$\rm \omega^{'} \,[\degr]$     		&$56			\pm 14$\\
$T^{'}$~[HJD]                 		&$2456290		\pm 50 $\\
$\rm P_{12}$~[year]       			&$12.6			\pm 0.3 $\\
$A$~[day]                	 		&$0.0030		\pm 0.0002$ \\
$\rm f(m_{3})~[M_{\odot}]$			&$0.0009		\pm 0.0002$ \\
$\rm M_{3}~[M_{\odot}]$   			&$0.15			\pm 0.01$ \\
\tableline
\end{tabular}
\tablecomments{
\begin{itemize}
\itemsep-0.5em
\item $A$ --- the semi-amplitude of the light-time effect.
\item $T^{'}$ --- the periastron passage.
\item $\rm P_{12}$ --- the period of the hypothetical third body.
\item $\rm a^{'}_{12}\sin i^{'}$ --- the semi-major axis.
\item $\rm e^{'}$ --- the eccentricity.
\item $\rm \omega^{'}$ --- the longitude of the periastron passage of the
  orbit of the eclipsing pair around the mass center of the system.
\item $  \rm f(m_{3}) $ --- the mass function for the hypothetical third body.
\item $ \rm M_{3}$ --- the mass of the hypothetical third body.
\end{itemize}}
\end{footnotesize}
\end{table}

\end{document}